\documentclass[fleqn,twoside]{article}
\usepackage{espcrc2}

\usepackage{graphicx}
\usepackage[figuresright]{rotating}

\newcommand{\AmS}{{\protect\the\textfont2
  A\kern-.1667em\lower.5ex\hbox{M}\kern-.125emS}}

\title{The CDF Silicon Vertex Trigger\thanks{Talk presented at the 9th
Pisa Meeting on Advanced Detectors, La Biodola, Isola d'Elba, May
25-31, 2003.}}

\author{
Bill~Ashmanskas\address[Argonne]
{Argonne National Laboratory, Argonne, IL 60439,
USA}\thanks{Corresponding author.  E-mail:
\texttt{wja@hep.anl.gov}}, 
A.~Barchiesi\address[Roma]
{INFN, Sezione di Roma I and University La Sapienza, I-00173 Roma, Italy},
A.~Bardi\address[Pisa]
{INFN,  University and Scuola Normale Superiore of Pisa, I-56100 Pisa, Italy},
M.~Bari\address[Trieste]
{INFN, Sezione di Trieste,  I-34012 Trieste, Italy},
M.~Baumgart\address[Chicago]
{Enrico Fermi Institute, University of Chicago, Chicago, IL 60637, USA},
S.~Belforte\addressmark[Trieste],
J.~Berryhill\addressmark[Chicago],
M.~Bogdan\addressmark[Chicago],
R.~Carosi\addressmark[Pisa],
A.~Cerri\address[LBL]
{Lawrence Berkeley National Laboratory, Berkeley, CA 94720, USA},
G.~Chlachidze\address[Dubna]
{Joint Institute for Nuclear Research, Dubna, Russia},
R.~Culbertson\addressmark[Chicago],
M.~Dell'Orso\addressmark[Pisa],
S.~Donati\addressmark[Pisa],
I.~Fiori\address[Padova]
{University of Padova and INFN, Sezione di Padova, I-35031 Padova, Italy},
H.~Frisch\addressmark[Chicago],
S.~Galeotti\addressmark[Pisa],
P.~Giannetti\addressmark[Pisa],
V.~Glagolev\addressmark[Dubna],
A.~Leger\address[Geneve]
{University of Geneva, CH-122, Geneva 4, Switzerland},
Y.~Liu\addressmark[Geneve],
T.~Maruyama\addressmark[Chicago],
E.~Meschi\addressmark[Pisa],
L.~Moneta\addressmark[Geneve],
F.~Morsani\addressmark[Pisa],
T.~Nakaya\addressmark[Chicago],
G.~Punzi\addressmark[Pisa],
M.~Rescigno\addressmark[Roma],
L.~Ristori\addressmark[Pisa],
H.~Sanders\addressmark[Chicago],
S.~Sarkar\addressmark[Roma],
A.~Semenov\addressmark[Dubna],
M.~Shochet\addressmark[Chicago],
T.~Speer\addressmark[Geneve],
F.~Spinella\addressmark[Pisa],
H.~Vataga\addressmark[Pisa],
X.~Wu\addressmark[Geneve],
U.K.~Yang\addressmark[Chicago],
L.~Zanello\addressmark[Roma],
A.M.~Zanetti\addressmark[Trieste]
\centerline{(for the CDF-II Collaboration)}
}

\begin{document}

\begin{abstract}
The CDF experiment's Silicon Vertex Trigger is a system of 150 custom
9U VME boards that reconstructs axial tracks in the CDF silicon strip
detector in a 15~$\mu$sec pipeline.  SVT's 35~$\mu$m impact parameter
resolution enables CDF's Level 2 trigger to distinguish primary and
secondary particles, and hence to collect large samples of hadronic
bottom and charm decays.  We review some of SVT's key design features.
Speed is achieved with custom VLSI pattern recognition, linearized
track fitting, pipelining, and parallel processing.  Testing and
reliability are aided by built-in logic state analysis and test-data
sourcing at each board's input and output, a common inter-board data
link, and a universal ``Merger'' board for data fan-in/fan-out.  Speed
and adaptability are enhanced by use of modern FPGAs.
\vspace{1pc}
\end{abstract}

\maketitle

\section{CDF TRIGGER OVERVIEW}

The recently upgraded Collider Detector at Fermilab (CDF)
experiment~\cite{cdf} pursues a broad physics program at Fermilab's
Tevatron proton-antiproton collider, comprising topics as diverse as
top quark production and charmed meson decay.  In the present Tevatron
data-taking period (``Run 2''), the c.m. energy is
$\sqrt{s}=1.96$~TeV, the bunch-crossing interval is 396~ns (with a
possible upgrade to 132~ns for high luminosity), and peak luminosities
are $0.5\times10^{32}~{\rm cm}^{-2}{\rm s}^{-1}$ to date and climbing
toward a goal of $2\times10^{32}~{\rm cm}^{-2}{\rm s}^{-1}$.  CDF's
new drift chamber~\cite{ambrose} and silicon detector~\cite{lester} are
discussed in detail elsewhere in these proceedings.

One challenge for a hadron collider experiment is to extract signals
of interest efficiently from much larger backgrounds.  To illustrate
the orders of magnitude, the total inelastic cross-section at the
Tevatron is about 50~mb, while the $b$-quark cross-section within
CDF's acceptance (transverse momentum $p_T>6$~GeV, rapidity $|y|<1$)
is about 10~$\mu$b, and the $t$-quark cross-section is about 5~pb.  At
luminosities above $0.35\times10^{32}~{\rm cm}^{-2}{\rm s}^{-1}$, the
mean number of interactions per beam crossing exceeds 1.  Reducing the
1.7~MHz beam-crossing rate to CDF's 70~Hz DAQ output rate implies a
trigger rejection of 25000.

Good background rejection in the trigger requires fast identification
of distinctive signal signatures.  In the CDF trigger, many important
signatures exploit fast charged-particle track reconstruction in the
bending plane of the spectrometer, transverse to the beam axis.  The
trigger matches drift chamber tracks with EM calorimeter showers, muon
chamber stubs, and silicon detector data, respectively, to identify
electrons, muons, and $b$ and $c$ daughters.

CDF uses a three-level trigger.  On each beam crossing (396 or
132~ns), the entire front end digitizes (silicon samples and holds).
A 5.5~$\mu$s pipeline of programmable logic forms axial drift chamber
tracks and can match these with calorimeter and muon-chamber data.  On
Level~1 accept, front-end boards store the event to one of four
buffers (silicon digitizes and transmits to the silicon trigger and
event builder).  Level~2 processing, with about 30~$\mu$s latency,
adds fast silicon tracking, calorimeter clustering, and EM calorimeter
shower-max data.  The final Level~2 decision is made in software on a
single-board computer, so a wider range of thresholds and derived
quantities is possible (e.g. transverse mass of muon track pairs),
even for information that is in principle available at Level~1.  On
Level~2 accept, front-end VME crates transmit to the event builder.
At Level~3, a farm of 250 commodity PCs runs full event
reconstruction.  This is the first stage at which three-dimensional
tracks (e.g. for invariant mass calculation) are available.  Events
passing Level~3 are written to disk.

While some optimization remains to be done, the maximum output at
L1/L2/L3 is approximately 35000/350/70 Hz.  Each of these rates is an
order of magnitude higher than in CDF's 1992-96 running period.  In
addition, drift chamber tracking has moved from L2 to L1, and silicon
tracking has moved from offline to L2.  These three changes allow CDF
to collect large samples of fully hadronic bottom and charm decays, by
requiring two drift chamber tracks at L1, requiring each track to have
a significant (at least 120~$\mu$m) impact parameter at L2, and
performing full software tracking at L3 to confirm the hardware
tracking.  The samples made possible by CDF's front-end, trigger, and
DAQ upgrades have yielded novel physics results~\cite{ivan} at an
early stage of Run~2.

CDF's Level~1 drift chamber hardware track processor, XFT~\cite{xft},
is a cornerstone of the CDF trigger.  For every bunch crossing, with
1.9~$\mu$s latency, it finds tracks of $p_T>1.5$~GeV with 96\%
efficiency.  XFT obtains coarse hit data (two time bins) from each
axial drift chamber wire, finds line segments in the 12 measurement
layers of each axial superlayer, then links segments from these four
superlayers to form track candidates.  XFT's resolutions,
$\sigma(\frac{1}{p_T})=1.7\%/{\rm GeV}$ and $\sigma(\phi_0)=5$~mrad,
are only about a factor of 10 coarser than those of the offline
reconstruction.

\section{SVT TRACK PROCESSING}

For each event passing Level~1, the Silicon Vertex Trigger
(SVT)~\cite{belfo,svttdr,luciano} swims each XFT track into the silicon
detector, associates silicon hit data from four detector planes, and
produces a transverse impact parameter measurement of 35~$\mu$m
resolution (50~$\mu$m when convoluted with the beam spot) with a mean
latency of 24~$\mu$s, 9~$\mu$s of which are spent waiting for the
first silicon data.  SVT's impact parameter resolution for
$p_T\approx2$~GeV is comparable to that of offline tracks that do not
use Layer~00 (mounted on the beam pipe), which is not yet available in
SVT.

For fiducial offline muon tracks from $J/\psi$ decay, having
$p_T>1.5$~GeV and hits in the four silicon planes used by SVT,
measured SVT efficiency is 85\%.  The most suitable definition of
efficiency in a given context depends on what one aims to optimize:
restricting the denominator to $p_T>2$~GeV increases the efficiency to
90\%, while relaxing the requirements on which layers contain offline
silicon hits reduces the efficiency to 70\%, and looser fiducial
requirements reduce the efficiency further; the ultimate denominator
for SVT would be all XFT-matched offline silicon tracks that are
useful for physics analysis.

SVT is a system of 150 custom 9U VME boards containing FPGAs, RAMs,
FIFOs, and one ASIC design.  CPUs are used only for initialization and
monitoring.  SVT's input comprises 144 optical fibers, 1~Gbit/s each,
and one 0.2~Mbit/s LVDS cable; its output is one 0.7~Mbit/s LVDS
cable.

Three key features allow SVT to carry out in 15~$\mu$s a silicon track
reconstruction that typically requires ${\cal O}(0.1~{\rm s})$ in
software: a highly parallel/pipelined architecture, custom VLSI
pattern recognition, and a linear track fit in fast FPGAs.

The silicon detector's modular, symmetric geometry lends itself to
parallel processing.  SVT's first stage, converting a sparsified list
of channel numbers and pulse heights into charge-weighted hit
centroids, processes $12\times6\times5$ (azimuthal $\times$
longitudinal $\times$ radial) silicon planes in 360 identical FPGAs.
The overall structure of SVT reflects the detector's 12-fold azimuthal
symmetry.  Each $30^\circ$ azimuthal slice is processed in its own
asynchronous, data-driven pipeline that first computes hit centroids,
then finds coincidences to form track candidates, then fits the
silicon hits and drift chamber track for each candidate to extract
circle parameters and a goodness of fit.

In SVT's usual configuration, a track candidate requires a coincidence
of an XFT track and hits in a specified four (out of five available)
silicon layers.  To define a coincidence, each detector plane is
divided into bins of programmable width, typically 250-700~$\mu$m, and
XFT tracks are swum to the outer radius of the silicon detector and
binned with 3~mm typical width.  For each $30^\circ$ slice, the set of
32K most probable coincidences (``patterns'') is computed offline in a
Monte Carlo program and loaded into 256 custom VLSI associative memory
(AM) chips.  For every event, each binned hit is presented in parallel
to the 256 AM chips, and the hit mask for each of the 128 patterns per
chip is accumulated in parallel.  When the last hit has been read, a
priority encoder enumerates the patterns for which all five layers
have a matching hit.  The processing time is thus linear in the total
number of hits in each slice and linear in the number of {\em matched}
patterns.

There is no exact linear relationship between the transverse
parameters $c$, $\phi$, $d$ of a track in a solenoidal field and the
coordinates at which the track meets a set of flat detector planes:
the coordinates are more closely linear in $\frac{c}{\cos^3\phi}$,
$\tan\phi$, and $\frac{d}{\cos\phi}$.  But for $p_T>2$~GeV, $|d|<1$~mm,
$|\phi|<15^\circ$, a linear fit biases $d$ by at most a few percent.
By linear regression to Monte Carlo data, we derive the $3\times6$
coefficients $\mathbf V$ and 3 intercepts $\vec{p}_0$ relating
$\vec{p} = (c,\phi,d)$ to the vector $\vec{x}$ of $c_{\rm XFT}$,
$\phi_{\rm XFT}$, and four silicon hits: $\vec{p} = \vec{p}_0 +
{\mathbf V}\cdot\vec{x}$.  The same regression produces coefficients
$\mathbf C$ and intercepts $\vec{\chi}_0$, corresponding to the fit's
3 degrees of freedom, with which we calculate constraints $\vec{\chi}
= \vec{\chi_0} + {\mathbf C}\cdot\vec{x}$ and the usual $\chi^2 =
|\vec{\chi}|^2$.  In the start-of-run download, we precompute
$\vec{p}$ and $\vec{\chi}$ for the coordinates at the edge of each
pattern and store them in flash memory.  Using each candidate's
pattern ID as a hint, the fitter board computes corrections to
$\vec{p}$ and $\vec{\chi}$ with respect to the pattern edge, using
8-bit multiplication in 6 parallel FPGAs, in 250~ns per fitted track.
Tracks passing programmable goodness-of-fit cuts propagate downstream.

\section{SVT DIAGNOSTIC FEATURES}

An SVT whose processing time, resolution, or inefficiency were 20-30\%
larger would still have enabled novel physics results at CDF.  But an
SVT that could not be commissioned quickly or operated reliably would
have been a failure.  Several design features of SVT contributed to
its rapid commissioning and reliable operation.

The essence of SVT's component-based architecture is captured by the
SVT cable and the SVT Merger board.  Nearly all SVT internal
data---hit centroids, drift chamber tracks, pattern IDs, track
candidates, and fitted SVT tracks---travel as LVDS signals on common
26-conductor-pair cables carrying data bits, a data strobe, a
flow-control signal, and a ground pair.  The data are variable-length
packets of 21-bit words, plus end-packet and end-event bits.  Data
fan-in and fan-out are performed inside FPGAs, not on backplanes, by a
universal Merger board that concatenates event data for up to four SVT
cable inputs and provides two SVT cable outputs.  Every fan-in stage
compares event IDs for its sources and can drive a backplane error
line on mismatch.  A parity bit for each cable-event provides a basic
check of data integrity.  It is illustrative of SVT's design strategy
that the SVT cable and Merger board were prototyped and tested {\em
before} the boards to cluster hits, find and fit tracks, etc.

The Merger board is reminiscent of the fan-in/fan-out modules found in
NIM trigger electronics, and lends itself to the same kind of
inventive ad-hoc cabling for producing quick results in test stands
and during system commissioning.

On each end of every SVT cable is a circular memory buffer that
records---as a logic state analyzer---the last $10^5$ words sent or
received on that cable.  Comparing a sender's output buffer with a
receiver's input buffer checks data transmission.  Comparing a board's
input and output with emulation software checks data processing.  The
memories also serve as sources and sinks of test patterns for testing
single boards, a small chain of boards, a slice of SVT, SVT as a
standalone system, or the data paths to SVT's external sources and
sink.  The buffers can be frozen and read by monitoring software
parasitically during data-taking, and all of SVT's buffers can be
frozen together, via backplane signals, when any board detects an
error condition, such as invalid data.

By polling SVT's circular memories during beam running, large samples
of track and hit data, pattern IDs, etc.---unbiased by L2 or L3
trigger decisions---are sampled and statistically analyzed to monitor
data quality.  A beam-finding program monitors $10^7$ tracks per hour,
fitting and reporting to the accelerator control network an updated
Tevatron beamline fit every 30 seconds.  The beam fit is also written
to the DAQ event record and used to correct in-situ every SVT track's
impact parameter for the sinusoidal bias vs $\phi$ resulting from the
beamline's offset from the detector origin, so that the trigger is
immune to modest beam offsets.

The flexibility of FPGAs has been exploited throughout SVT, enabling
SVT to adapt to unforseen circumstances when commissioning the
detector and trigger as a whole.  Later boards had the benefit of more
flexible programmable chips.  In particular, the board that subtracts
the beam offset track-by-track---the last SVT board to be
built---illustrates well both the utility of modern FPGAs and the
virtue of a component-based architecture.  It was designed as a
clean-up board, beyond the SVT baseline, to ensure that at most one
SVT track is output per XFT track.  SVT's modularity allowed this
final processing stage to be added seamlessly.  Progress in FPGA
technology allowed the board to consist essentially of input circuitry
+ large FPGA + output circuitry.  With this design, it was
straightforward to adapt the board to subtract a sinusoidal beam
offset---which proved more convenient than the baseline plan to steer
the Tevatron beam.  This clean-up board has found even further uses,
such as recording SVT's event-by-event processing time into the DAQ
event record and online monitor.

In conclusion, the Silicon Vertex Trigger has been commissioned and
operated successfully for CDF's first year of Run~2 physics data.
Among the key reasons for this system's success are its modular
architecture and its ability to sink and source test data at a wide
range of pipeline stages, both in tests and during beam runs.  SVT's
flexibility and diagnostic features were particularly valuable during
the CDF commissioning period.

\end{document}